\documentclass[cits]{PoS}

\usepackage{amsmath,amsfonts,amssymb,slashed,dsfont}
\usepackage{booktabs}

\newcommand{\Order}{\mathcal{O}}

\newcommand{\mpi}{M_\pi}

\newcommand{\Fpi}{F_\pi}

\newcommand{\beq}{\begin{equation}}
\newcommand{\eeq}{\end{equation}}
\newcommand{\muu}{m_\text{u}}
\newcommand{\md}{m_\text{d}}

\newcommand{\diff}{\text{d}}
\newcommand{\tm}{t_\text{m}}
\newcommand{\ff}{\boldsymbol{f}}
\newcommand{\tpi}{t_\pi}
\newcommand{\tK}{t_K}

\providecommand{\MeV}{\,\text{MeV}}
\providecommand{\GeV}{\,\text{GeV}}

\renewcommand{\Im}{\text{Im}\,}

\title{Improved dispersive analysis of the scalar form factor of the nucleon}

\ShortTitle{Improved dispersive analysis of the scalar form factor of the nucleon}

\author{\speaker{Martin Hoferichter}, $^{ab}$ Christoph Ditsche, $^a$ Bastian Kubis, $^a$ and Ulf-G.~Mei\ss ner$^{ac}$\\
       \llap{$^{a}$} Helmholtz-Institut f\"{u}r Strahlen- und Kernphysik (Theorie), 
Bethe Center for Theoretical Physics, Universit\"at Bonn, D-53115 Bonn, 
Germany\\
\llap{$^{b}$} Albert Einstein Center for Fundamental Physics, Institute for Theoretical Physics,
	    Universit\"at Bern, CH-3012 Bern, Switzerland\\
\llap{$^{c}$} Forschungszentrum J\"ulich, Institut f\"ur Kernphysik, 
J\"ulich Center for Hadron Physics  and Institute for Advanced Simulation,  D-52425 J\"ulich, Germany\\
E-mail: \email{hoferichter@itp.unibe.ch}}

\abstract{We present a coupled system of integral equations for the $\pi\pi\to\bar N N$ and $\bar K K\to\bar N N$ $S$-waves derived from Roy--Steiner equations for pion--nucleon scattering. We discuss the solution of the corresponding two-channel Muskhelishvili--Omn\`es problem and apply the results to a dispersive analysis of the scalar form factor of the nucleon fully including $\bar K K$ intermediate states. In particular, we determine the corrections $\Delta_\sigma$ and $\Delta_D$, which are needed for the extraction of the pion--nucleon $\sigma$ term from $\pi N$ scattering, and show that the difference $\Delta_D-\Delta_\sigma=(-1.8\pm 0.2)\MeV$ is insensitive to the input $\pi N$ parameters.}

\FullConference{The 7th International Workshop on Chiral Dynamics,\\
		August 6 -10, 2012\\
		Jefferson Lab, Newport News, Virginia, USA}

\begin{document}

\section{Introduction}

The pion--nucleon $\sigma$ term $\sigma_{\pi N}$ measures the contribution of the light quarks to the nucleon mass $m$, and is directly related to the form factor of the scalar current 
\beq
\sigma(t)=\frac{1}{2m}\langle N(p')|\hat m (\bar u u+ \bar d d)|N(p)\rangle,\qquad t=(p'-p)^2,\qquad \hat m=\frac{\muu+\md}{2},
\eeq
at vanishing momentum transfer $\sigma(0)=\sigma_{\pi N}$. The standard procedure for its extraction from pion--nucleon ($\pi N$) scattering relies on the low-energy theorem~\cite{ChengDashen,BPP}
\beq
\label{LET}
\Fpi^2\bar D^+\big(s=m^2,t=2\mpi^2\big)=\sigma\big(2\mpi^2\big)+\Delta_\text{R},
\eeq
which relates the Born-term-subtracted isoscalar $\pi N$ scattering amplitude at the Cheng--Dashen point $\bar D^+(s=m^2,t=2\mpi^2)$ to the scalar form factor evaluated at $2\mpi^2$. The remainder $\Delta_\text{R}$ is free of chiral logarithms at full one-loop order in chiral perturbation theory (ChPT)~\cite{BKM96,BL01}, and has been estimated as~\cite{BKM96}
\beq
|\Delta_\text{R}|\lesssim 2\MeV.
\eeq
Rewriting~\eqref{LET} in terms of
\beq
\Delta_D=\Fpi^2\Big\{\bar D^+\big(s=m^2,t=2\mpi^2\big)-d_{00}^+-2\mpi^2 d_{01}^+\Big\},\qquad 
\Delta_\sigma=\sigma\big(2\mpi^2\big)-\sigma_{\pi N},
\eeq
the extraction of the $\sigma$ term reduces to the determination of the subthreshold parameters $d_{00}^+$ and $d_{01}^+$ as well as the combination
$\Delta_D-\Delta_\sigma-\Delta_\text{R}$. The first two corrections can be calculated using a dispersive approach~\cite{GLS90_FF}
\beq
\Delta_D-\Delta_\sigma=(-3.3\pm 0.2)\MeV,
\eeq
where the error only covers the uncertainties in the $\pi\pi$ phase shifts available at that time. Here, we update the determination of $\Delta_D$ and $\Delta_\sigma$ using modern $\pi\pi$ phases, fully including $\bar K K$ intermediate states, and carefully studying the dependence of the results on $\pi N$ subthreshold parameters as well as the $\pi N$ coupling constant.

\section{Scalar pion and kaon form factors}

We first consider the case of the scalar pion and kaon form factors $F_\pi^S(t)$ and $F_K^S(t)$, which serve both to illustrate the method and as input for the scalar form factor of the nucleon. Unitarity in the $\pi\pi/\bar K K$ system intertwines both form factors according to~\cite{DGL90} 
\beq
\label{meson_unitarity}
\Im \boldsymbol{F}^S(t)=\big(T(t)\big)^*\Sigma(t) \boldsymbol{F}^S(t),\qquad
\boldsymbol{F}^S(t)=\begin{pmatrix}F_\pi^S(t)\\\frac{2}{\sqrt{3}}F_K^S(t)\end{pmatrix},
\eeq
with the phase-space factor
\beq
\Sigma(t)=\text{diag}\Big(\sigma^\pi_t\theta\big(t-\tpi\big),\sigma^K_t\theta\big(t-\tK\big)\Big),\qquad 
\sigma^i_t=\sqrt{1-\frac{t_i}{t}},\qquad t_i=4M_i^2\qquad i\in\{\pi,K\},
\eeq
and the $T$-matrix
\beq
T(t)=\begin{pmatrix}
      \frac{\eta^0_0(t) e^{2i\delta^0_0(t)}-1}{2i\sigma_t^\pi} & |g(t)|e^{i\psi^0_0(t)}\\ 
  |g(t)|e^{i\psi^0_0(t)} & \frac{\eta^0_0(t) e^{2i(\psi^0_0(t)-\delta^0_0(t))}-1}{2i\sigma_t^K}
\end{pmatrix},
\eeq
expressed in terms of the $\pi \pi$ and $\pi\pi\to\bar K K$ phase shifts $\delta^0_0$ and $\psi^0_0$ as well as the inelasticity parameter $\eta^0_0=\sqrt{1-4\sigma^\pi_t\sigma^K_t|g(t)|^2\theta\big(t-\tpi\big)}$. The two-channel Muskhelishvili--Omn\`es (MO) problem~\cite{Muskhelishvili,Omnes} defined by the unitarity relation~\eqref{meson_unitarity} permits two linearly independent solutions $\boldsymbol{\Omega}_1$, $\boldsymbol{\Omega}_2$~\cite{Muskhelishvili}, which may be combined in the Omn\`es matrix $\Omega(t)$. In general, there is no analytical solution for $\Omega(t)$, we follow here the discretization method of~\cite{Moussallam99} for its numerical calculation.  

\begin{figure}
\centering
\includegraphics[height=0.244\textheight]{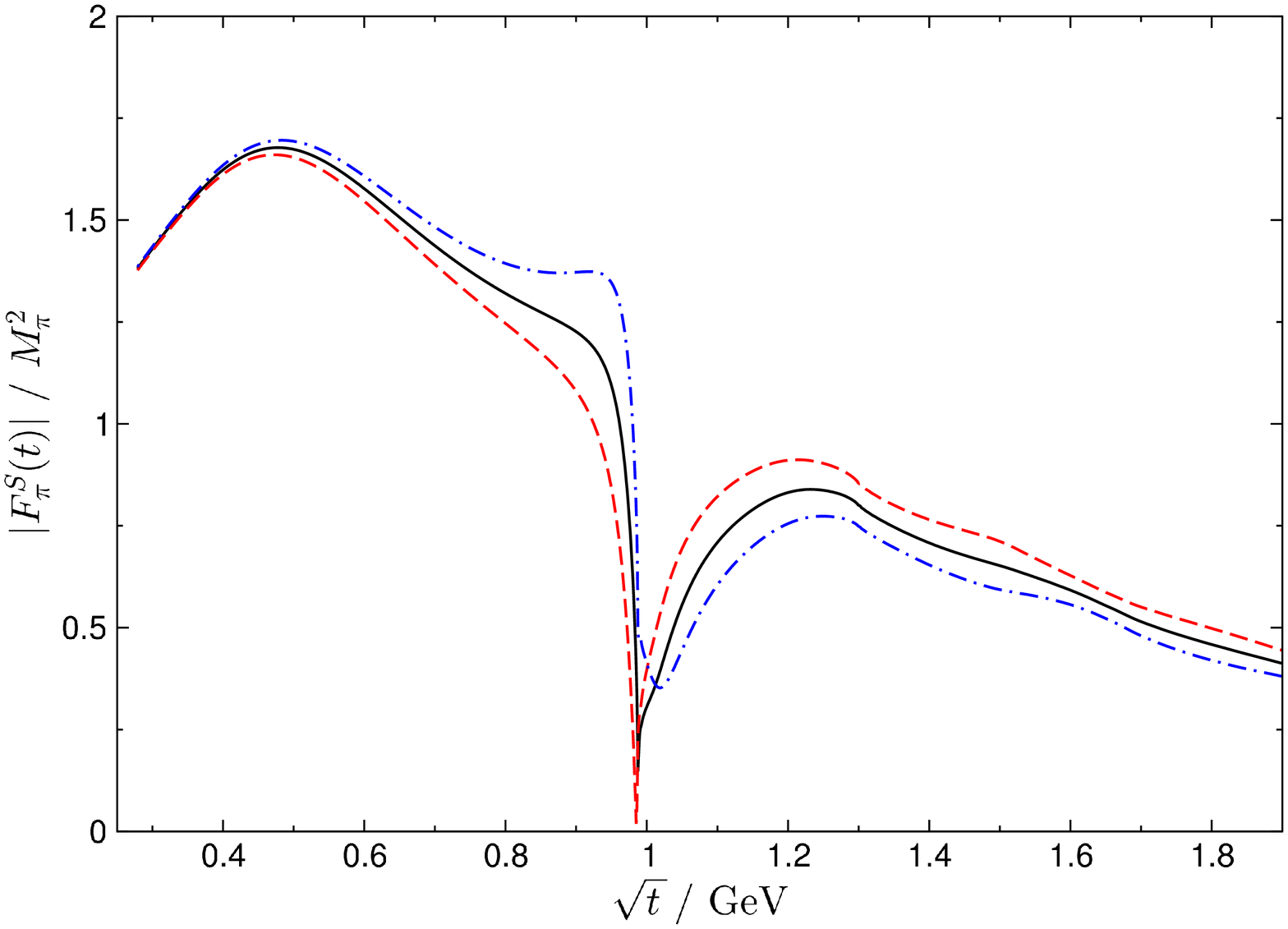}
\includegraphics[height=0.244\textheight]{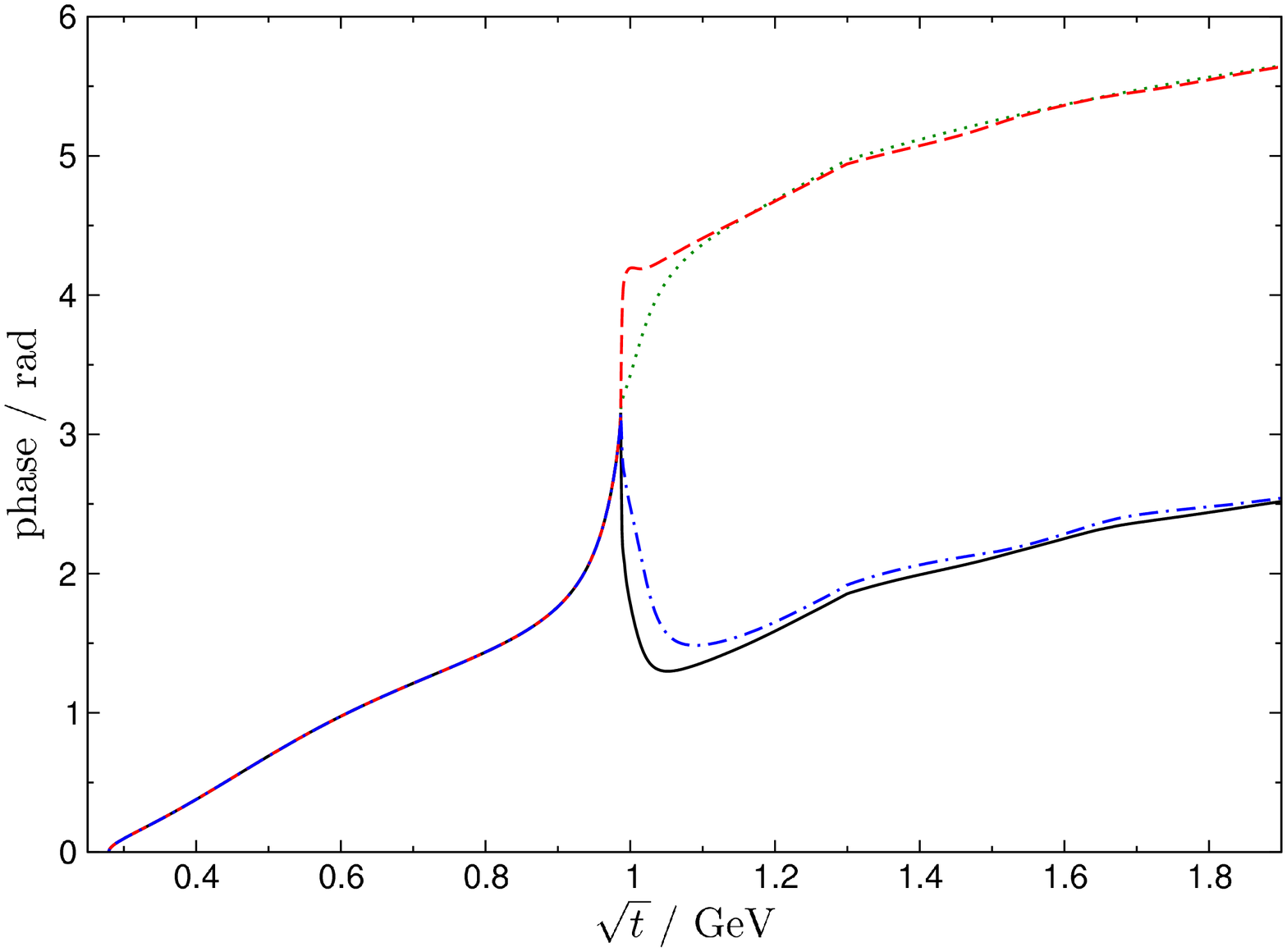}
\caption{Modulus and phase of the scalar pion form factor. The solid, dashed, and dot-dashed lines refer to $F^S_K(0)=\mpi^2/2$, $0.4\,\mpi^2$, and $0.6\,\mpi^2$. The phase of $F^S_\pi(t)$ is compared to $\delta^0_0$, as indicated by the dotted line.}
\label{fig:SFF}
\end{figure}

Since the form factors are devoid of a left-hand cut, they are related directly to the solutions of the MO problem with coefficients determined by $F_\pi^S(0)$ and $F_K^S(0)$~\cite{DGL90}. Using ChPT at $\Order(p^4)$ and the low-energy constants from~\cite{FLAG} we find
\beq
F_\pi^S(0)=(0.984\pm 0.006)\mpi^2,\qquad F_K^S(0)=(0.4\ldots 0.6)\mpi^2,
\eeq
which, together with $\delta^0_0$ and $\eta^0_0$ from an extended Roy-equation analysis of $\pi\pi$ scattering~\cite{CCL:Regge+PWA}, $\psi^0_0$ from partial-wave analyses~\cite{CohenEtkin}, and $|g(t)|$ from a Roy--Steiner (RS) analysis of $\pi K$ scattering~\cite{BDM04}, yield the results for $F_\pi^S(t)$ depicted in Fig.~\ref{fig:SFF}. The strong dependence of $F_\pi^S(t)$ near $\tK$ on $F^S_K(0)$ attests to the inherent two-channel nature of the problem and implies that an effective single-channel description in terms of the phase of $F_\pi^S(t)$ only works for sufficiently large $F_K^S(0)$. 

\section{From Roy--Steiner equations to the scalar form factor}
\label{sec:RS_SFF}

Unitarity couples the $\pi\pi\to\bar N N$ and $\bar K K\to\bar N N$ $S$-waves $f^0_+(t)$ and $h^0_+(t)$ analogously to~\eqref{meson_unitarity}
\beq
\Im \boldsymbol{f}(t)=\big(T(t)\big)^*\Sigma(t) \boldsymbol{f}(t),\qquad 
\boldsymbol{f}(t)=\begin{pmatrix}f^0_+(t)\\\frac{2}{\sqrt{3}}h^0_+(t)\end{pmatrix},
\eeq
but due to the presence of the left-hand cut the solution of the corresponding MO problem involves inhomogeneous contributions,
which may be derived from RS equations, cf.~\cite{BDM04,DHKM,HPS}. Generically, the integral equation takes the form
\beq
\ff(t)=\boldsymbol{\Delta}(t)+(\boldsymbol{a}+\boldsymbol{b}t)(t-4m^2)+\frac{t^2(t-4m^2)}{\pi}\int\limits_{\tpi}^\infty\diff t'\frac{\Im\ff(t')}{t'^2(t'-4m^2)(t'-t)},
\eeq
where $\Delta(t)$ includes Born terms, $s$-channel integrals, and higher $t$-channel partial waves, while $\boldsymbol{a}$ and $\boldsymbol{b}$ subsume subthreshold parameters that emerge as subtraction constants. 
The main difficulty in the evaluation of the formal solution
\begin{align}
\ff(t)&=\boldsymbol{\Delta}(t)+(t-4m^2)\Omega(t)(\mathds{1}-t\,\dot\Omega(0))\boldsymbol{a}+t(t-4m^2)\Omega(t)\boldsymbol{b}\\
&-\frac{t^2(t-4m^2)}{\pi}\Omega(t)\int\limits_{\tpi}^{\tm}\diff t'\frac{\Im\Omega^{-1}(t')\boldsymbol{\Delta}(t')}{t'^2(t'-4m^2)(t'-t)}
+\frac{t^2(t-4m^2)}{\pi}\Omega(t)\int\limits_{\tm}^\infty\diff t'\frac{\Omega^{-1}(t')\Im \ff(t')}{t'^2(t'-4m^2)(t'-t)},\notag
\end{align}
concerns the construction of the Omn\`es matrix for a finite matching point $\tm$~\cite{DHKM}.

\begin{figure}
\centering
\includegraphics[width=0.493\textwidth]{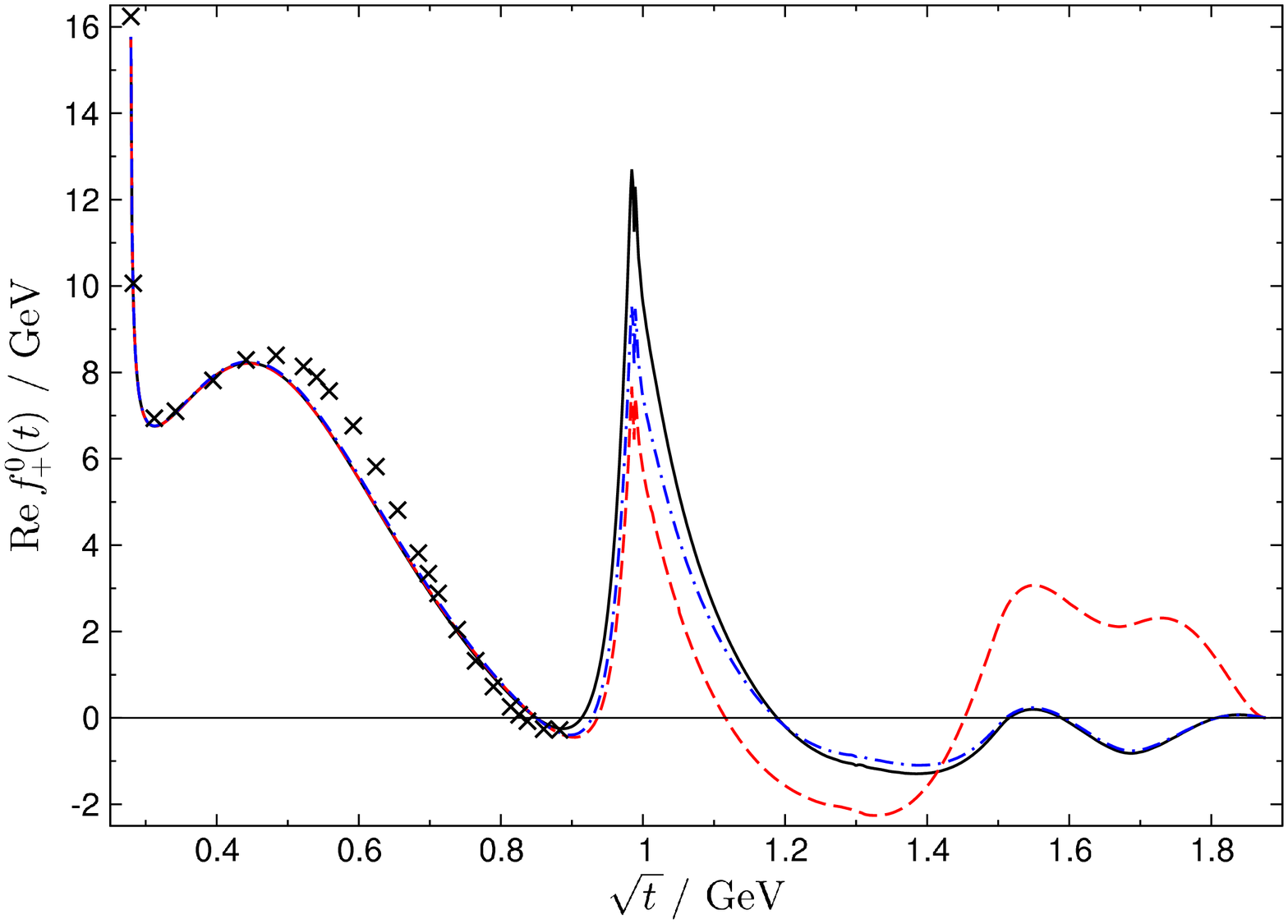}
\includegraphics[width=0.493\textwidth]{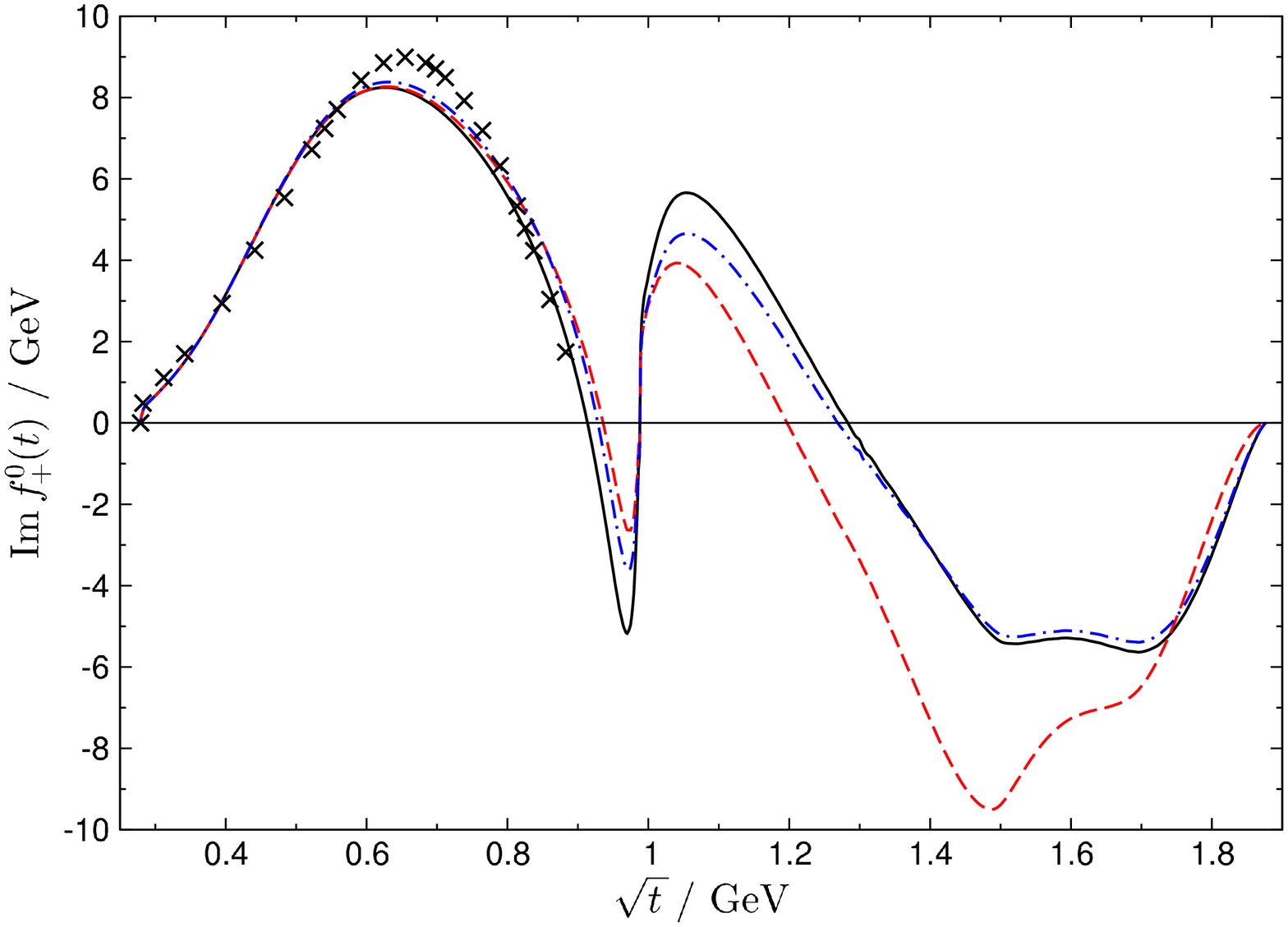}
\caption{Results for the real and imaginary part of $f^0_+(t)$. The solid, dashed, and dot-dashed lines refer to the input RS1, RS2, and RS3 as described in the main text. The black crosses indicate the results of~\cite{Hoehler}.}
\label{fig:f0p}
\end{figure}

In the numerical analysis we put $\Im \ff(t)=0$ above $\tm$, which we choose as $\tm=4m^2$ (thus exploiting a kinematical zero of $\ff(t)$), take the $\pi N$ and $KN$ $s$-channel partial waves from~\cite{Arndt_PWA}, and use the KH80 $\pi N$ coupling constant and subthreshold parameters as reference point~\cite{Hoehler}. In order to assess the uncertainties for higher energies we consider the following variants of the input: first, we keep the phase shifts $\delta^0_0$ and $\psi^0_0$ constant above $\sqrt{t_0}=1.3\GeV$ (``RS1''), where $4\pi$ intermediate states become important
and thus the two-channel approximation will break down, and second, guide both phase shifts smoothly to their asymptotic value of $2\pi$ as for the meson form factors (``RS2''). Finally, we amend RS1 in such a way that $\Delta_2(t)$, the $KN$ component of the inhomogeneity, is put to zero in order to assess the uncertainty in the $KN$ input (``RS3''). The corresponding results for $f^0_+(t)$ depicted in Fig.~\ref{fig:f0p} show that the largest uncertainty is induced by the high-energy phase shifts. 

\section{Results}

\begin{figure}
\centering
\includegraphics[width=0.493\textwidth]{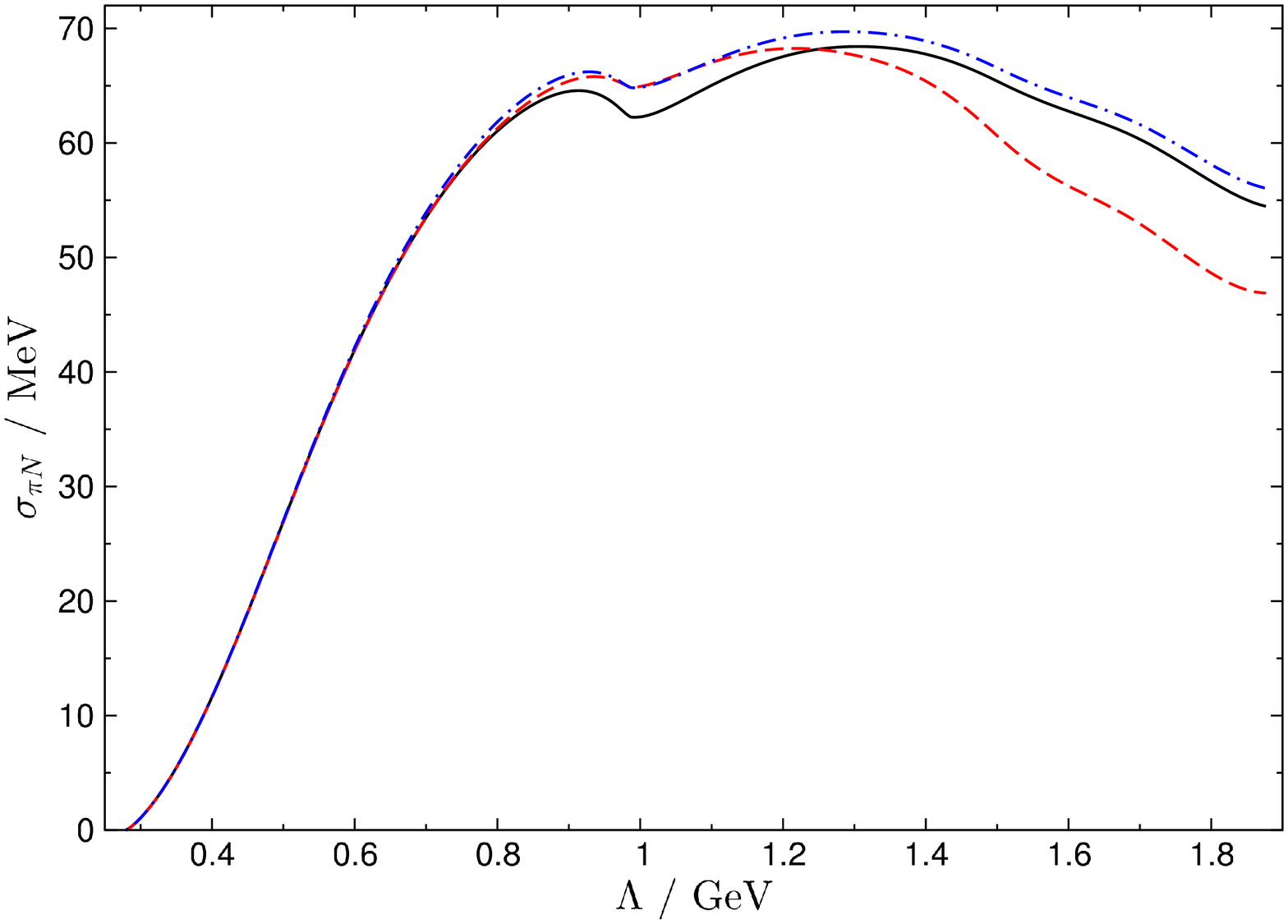}
\includegraphics[width=0.493\textwidth]{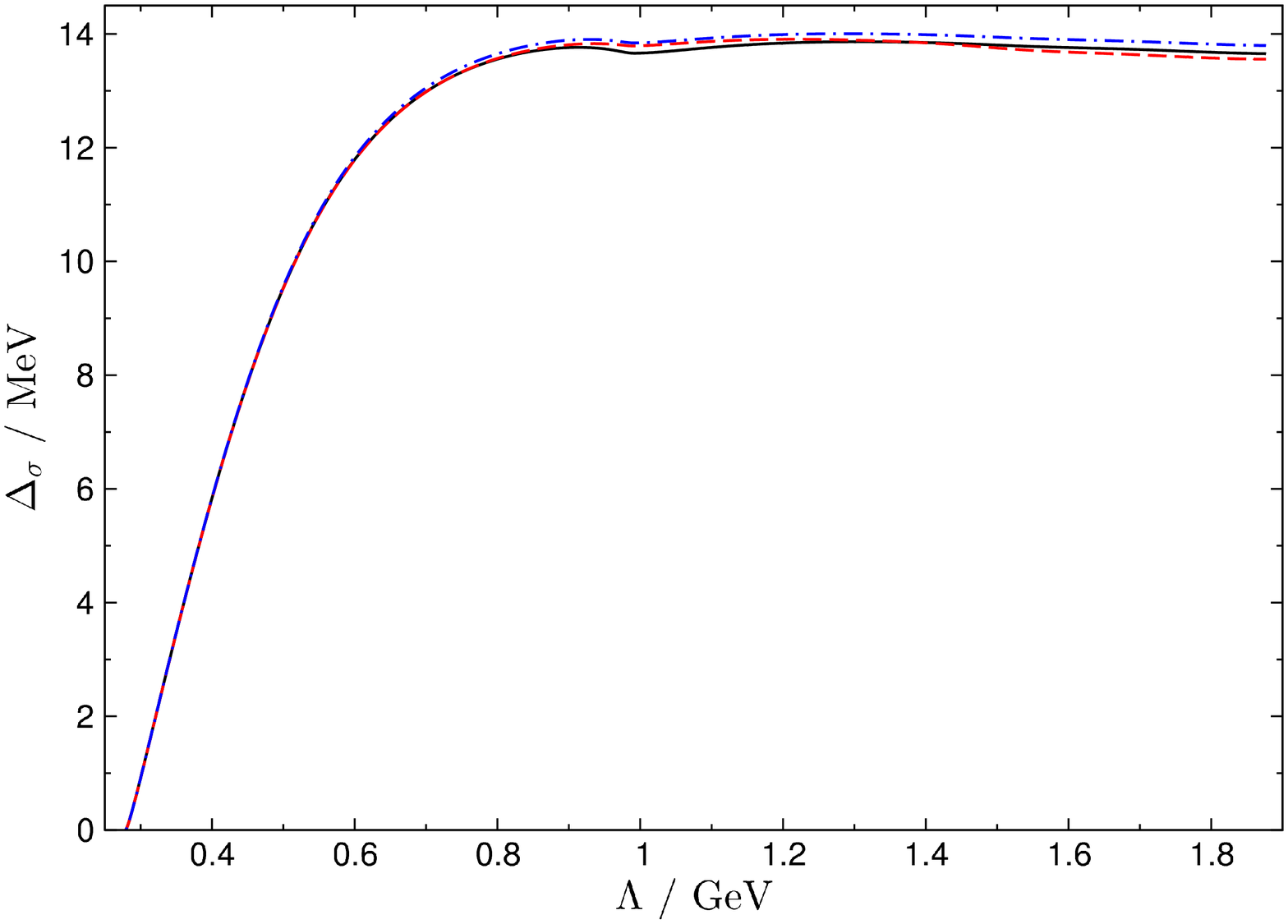}
\caption{$\sigma_{\pi N}$ and $\Delta_\sigma$ as a function of the integral cutoff $\Lambda$.}
\label{fig:sigma}
\end{figure}

The scalar form factor of the nucleon fulfills the unitarity relation
\beq
\Im \sigma(t)=\frac{2}{4m^2-t}\bigg\{\frac{3}{4}\sigma_t^\pi\big(F^S_\pi(t)\big)^*f^0_+(t)\theta\big(t-\tpi\big)
+\sigma_t^K\big(F^S_K(t)\big)^*h^0_+(t)\theta\big(t-\tK\big)\bigg\},
\eeq
so that, based on the results of the previous sections, the un- and once-subtracted dispersion relations
\beq
\sigma(t)=\frac{1}{\pi}\int\limits_{\tpi}^\infty\diff t'\frac{\Im\sigma(t')}{t'-t}=
\sigma_{\pi N}+\frac{t}{\pi}\int\limits_{\tpi}^\infty\diff t'\frac{\Im\sigma(t')}{t'(t'-t)}
\eeq
evaluated at $t=0$ and $t=2\mpi^2$ in principle determine $\sigma_{\pi N}$ and $\Delta_\sigma$ provided the two-channel approximation for the spectral function is sufficiently accurate in the energy range dominating the dispersive integral. Fig.~\ref{fig:sigma} shows that, while the dispersion relation converges too slowly for the $\sigma$ term itself, the result for $\Delta_\sigma$ becomes stable for $\Lambda\gtrsim 1\GeV$. Adding the uncertainties from the spectral function and the variation of the integral cutoff between $\Lambda=1.3\GeV$ and $\Lambda=2m$, we find
\begin{align}
\label{del_sig}
\Delta_\sigma&=(13.9\pm 0.3)\MeV\notag\\
&+ Z_1 \bigg(\frac{g^2}{4\pi}-14.28\bigg)+ Z_2\Big(d_{00}^+\,\mpi+1.46\Big)+
Z_3\Big(d_{01}^+\,\mpi^{3}-1.14\Big)+Z_4\Big(b_{00}^+\,\mpi^{3}+3.54\Big),\notag\\
Z_1&=0.36\MeV,\qquad Z_2= 0.57\MeV,\qquad Z_3= 12.0\MeV,\qquad Z_4=-0.81\MeV,
\end{align}
where we have made the dependence on the $\pi N$ parameters explicit (note that more modern determinations point to lower values of the $\pi N$ coupling constant around $g^2/4\pi\sim 13.7$~\cite{Nijmegen:1997,GWU:2006,piNcoupling:short+long}).

Similarly, the correction $\Delta_D$ follows from the $t$-channel expansion
\beq
\bar D^+(s=m^2,t)=d_{00}^++d_{01}^+ t-16 t^2\int\limits_{\tpi}^\infty\diff t'\frac{\Im f_+^0(t')}{t'^2(t'-4m^2)(t'-t)}
+\big\{ J\geq 2 \big\} + \big\{s\text{-channel integrals}\big\}
\eeq
evaluated at $t=2\mpi^2$, which gives
\begin{align}
 \Delta_D&=(12.1\pm 0.3)\MeV\notag\\
&+ \tilde Z_1 \bigg(\frac{g^2}{4\pi}-14.28\bigg)+ \tilde Z_2\Big(d_{00}^+\,\mpi+1.46\Big)+
\tilde Z_3\Big(d_{01}^+\,\mpi^{3}-1.14\Big)+\tilde Z_4\Big(b_{00}^+\,\mpi^{3}+3.54\Big),\notag\\
\tilde Z_1&=0.42\MeV,\qquad \tilde Z_2= 0.67\MeV,\qquad \tilde Z_3= 12.0\MeV,\qquad \tilde Z_4=-0.77\MeV.
\end{align}
Comparison with~\eqref{del_sig} shows that the dependence on the $\pi N$ parameters cancels nearly completely in the difference 
\beq
\Delta_D-\Delta_\sigma=(-1.8\pm 0.2)\MeV.
\eeq
This cancellation can be explained by the observation that the spectral function in both dispersion relations involves $f^0_+$ in a very similar manner, so that both integrals are equally affected by the dependence on the $\pi N$ parameters inherited from $f^0_+$. In the same way, part of the uncertainties discussed in Sect.~\ref{sec:RS_SFF} drop out, so that the final error estimate for $\Delta_D-\Delta_\sigma$ even decreases compared to the uncertainty in both corrections individually.

\end{document}